\documentclass[10pt]{article}

\usepackage{amsmath}
\usepackage{amssymb}

\usepackage[dvips,pdftex]{graphicx}

\usepackage{cite}

\usepackage{color} 
\usepackage{ulem}

\usepackage[labelfont=bf,labelsep=period,justification=raggedright]{caption}

\makeatletter
\renewcommand{\@biblabel}[1]{\quad#1.}
\makeatother

\date{}

\begin{document}

\begin{flushleft}
{\Large
\textbf{Assessing T cell clonal size distribution: a non-parametric approach}
}
\\
O.V. Bolkhovskaya$^{1}$, 
D.Yu. Zorin$^{2}$, 
M.V. Ivanchenko$^{2,\ast}$
\\
\bf{1} Faculty for Radiophysics, Department of Statistics, Lobachevsky State University of Nizhny Novgorod, Nizhny Novgorod, Russia
\\
\bf{2} Faculty for Computational Mathematics and Cybernetics, Department of Bioinformatics, Lobachevsky State University of Nizhny Novgorod, Nizhny Novgorod, Russia
\\
$\ast$ E-mail: ivanchenko.mv@gmail.com
\end{flushleft}

\section*{Abstract}
Clonal structure of the human peripheral T-cell repertoire is shaped by a number of homeostatic mechanisms, including antigen presentation, cytokine and cell regulation. Its accurate tuning leads to a remarkable ability to combat pathogens in all their variety, while systemic failures may lead to severe consequences like autoimmune diseases. Here we develop and make use of a non-parametric statistical approach to assess T cell clonal size distributions from recent next generation sequencing data. For  $41$ healthy individuals and a patient with ankylosing spondylitis, who undergone treatment, we invariably find power law scaling over several decades and for the first time calculate quantitatively meaningful values of decay exponent. It has proved to be much the same among healthy donors, significantly different for an autoimmune patient before the therapy, and converging towards a typical value afterwards. We discuss implications of the findings for theoretical understanding and mathematical modeling of adaptive immunity.

\section*{Introduction}

T lymphocytes are the key drivers of adaptive immune system, a powerful machinery able to detect, combat, and memorize pathogens in all their possible variety \cite{textbook}. Specific recognition of potentially harmful foreign peptides is achieved by the highly selective binding of T cell receptors (TCRs) to peptide-MHC complexes (p-MHC), mounted on the surface of specialized antigen-presenting cells. Required diversity of recognition arises due to an astronomic number of distinct molecular variants of TCRs, potentially emerging through random-like V(D)J recombination during cell development in thymus. In result of allelic exclusion, a T cell typically expresses a single TCR variant, and all its daughter cells have identical antigen recognition properties, constituting a clonotype \cite{Khor2002}. After massive positive and negative selection in thymus aimed to remove incapable or potentially autoimmune cells, about $5\%$ of {  thymocytes} finally enter periphery. Quantitative statistical assessment of TCR diversity is now becoming possible due to recent developments of next generation sequencing (NGS) techniques, capable of producing large TCR libraries. Current estimates claim about $10^{11}-10^{12}$ T cells compartmented in about $10^8$ clonotypes in healthy adult humans \cite{Arstila1999,Robins2009,Freeman2009,Warren2011}. 

Efficient and adequate performance of this ensemble requires fine tuning of T cell populations, which is achieved by a plethora of mechanisms: cytokine, antigen-presentation, and cell regulation \cite{Surth2008,Takada2009}. The nature of these mechanisms implies that competition for cytokines controls the total number of T-cells, while competition for p-MHC binding sites adjusts the frequencies of individual clones. The latter involves a cascade of processes: proteasomal degradation of proteins, peptide delivery and presentation on MHC complexes, p-MHC - TCR binding events, epitope recognition and signaling, cell fate decision by regulatory networks. Since their thorough experimental quantification faces tremendous difficulties, it becomes appealing to assess at least the outcome. 

Attention is getting drawn to analyzing experimentally obtained clonal frequency (relative size) distributions, and there is a growing evidence that they are strongly non-Gaussian. Even the first sequencing results for human TCR repertoires (restricted to clonotypes responsive to a certain peptide, though) indicated that it has heavy tails that could follow a power law \cite{Naumov2003}.  Parametric approaches quite successfully employed several Poisson abundance model distributions for mice with limited TCR diversity \cite{Sepulveda2010,Rempala2011}. However, the studied TCR libraries contained only about a hundred of entries, which precluded a definite choice of the best model among the considered \cite{Sepulveda2010}. 

In contrast, the recently produced and publicly available NGS human TCR libraries contain $10^4 - 10^6$ distinct variants from {  41} reportedly healthy \cite{Warren2011,Britanova2014} and one autoimmune \cite{Mamedov2011,Bolotin2012,Britanova2012} adult donors, the latter screened before and after chemotherapy and autologous hematopoietic stem cell transfer (HSCT). It gives an opportunity to bring the quantitative analysis of the peripheral T-cell pool statistics on a qualitatively new level. 

It is worth noting, however, that quantitatively meaningful estimates of power law distributions demand careful and critical analysis \cite{Stumpf2012}. In particular, assessing clonotype distributions requires accurate treatment of statistical sampling error, especially significant for low-frequency clonotypes abundant in the T cell repertoire. Otherwise, the blind cutoff of that part of repertoire would seriously restrict the volume of data and span coverage, undermining statistical significance of the result. This has proved to be characteristic of the recent report on the ``fractal'' organization of the T cell repertoire, with convincing evidence of heavy tailed distributions but quite questionable numerics of power law fits \cite{Meier2013}.

The purpose of this paper is to conduct a comparative non-parametric analysis to estimate the T cell clonal frequency distributions  {  from a number of available human libraries} without any prior assumption on the functional form. We aim to determine whether reconstructed clonal frequency distributions follow some functional dependence, common between different healthy individuals, whether it changes in the autoimmune patient before and after treatment, whether there are differences in healthy and autoimmune repertoire distributions, if any. We reveal the power law scaling over several decades of distributions and quantify the exponents.    

\section*{Materials \& Methods}\label{MaM}

We analyze the sets of data coming from: (i) reportedly healthy donors: {  middle-aged} two male and one female subjects \cite{Warren2011}, {   $38$ subjects aged $9$ -- $90$ years \cite{Britanova2014} }, and (ii) a {  middle-aged} patient with ankylosing spondylitis who undergone chemotherapy and autologous HSCT, and was observed up to 25 months since the treatment \cite{Britanova2012} (see Table \ref{Table1} for details). The respective TCR libraries were build by a common workflow that is extraction of peripheral blood mononuclear cells, isolation of RNA, cDNA synthesis, PCR amplification and sequencing of TCR$\beta$ {  CDR3} region \cite{Warren2011,Mamedov2011,Bolotin2012,Britanova2012,Britanova2014}.  

It is essential that each step of {  profiling} is prone to experimental and statistical errors that depend on implementation details and equipment (see \cite{Warren2011,Bolotin2012,Venturi2012} for a detailed discussion). We trust the quality assessment and error correction for the reads that the authors of the TCR libraries had performed and make use of their final post-processed data. Following the common approach we associate different TCR variants with different T cell clonotypes. 

We focus our attention on the distribution of the multiply read TCR variants, since only they can be reliably distinguished from sequencing errors and allow a trustworthy frequency estimate. Instructively, even the most frequent clonotypes demonstrate significant variances of clonal frequencies in comparative deep sequencing of the same donor either within a week interval \cite{Warren2011} or in parallel draws \cite{Bolotin2012}. Therefore, the straightforward histogram analysis of the statistics of TCR reads is a questionable estimate of the clonotype frequency distribution, and it is essential to view the NGS repertoire profiling as a { $K$}-stage random sampling process, {  respective to sequencing steps}. {  Notably, the novel cDNA barcoding procedure establishes a one-to-one correspondence between a T cell in a sample and a sequenced TCR variant \cite{Britanova2014}, which allows for reducing the model to a single-step random sampling for their data.}

To incorporate statistical uncertainty in our analysis we construct a non-parametric kernel distribution estimator for the complementary cumulative frequency distribution  $\hat{F}(p^*)\equiv\mbox{Prob}(p\ge p^*)$ \cite{Rosenblatt1956,Parzen1962}:
\begin{equation}
\label{eq1}
\hat{F}_c(p)=\frac{1}{N_c}\sum\limits_{i=1}^{N_c} \hat{F}_{norm}(p-p_i),
\end{equation}
where $N_c$ is the number of {  identified} distinct sequences { (`distinct reads')}, $p_i$ is the frequency of each TCR variant, $\hat{F}_{norm}(\cdot)$ is the kernel function that is the complementary cumulative normal distribution with the zero mean and standard deviation 
\begin{equation}
\label{eq2}
\sigma_{p_i}=\left[\sum\limits_{k=1}^K\sigma_{p_i;k}^2\right]^{1/2}
\end{equation}
dependent on standard deviations $\sigma_{p_i;k}$ present at each of $K$ steps of repertoire profiling. 
Assuming that sampling a particular variant $i$ at step $k$ is a binomial process with probability $p_i$ we get
$\sigma_{p_i;k}=\sqrt\frac{p_i(1-p_i)}{N_k}$, where $N_k$ is the total number of samples at this step. {  Note, that as the expected value of a binomial process $\propto p_i$, the expected frequency of a TCR derivative product will remain $p_i$ at each step of sequencing.}

Obviously, the contribution of different steps to statistical errors is quite different (see Table \ref{Table1} for indicative sample sizes). Three major bottlenecks accumulating statistical error in our case are sampling T cells, synthesized cDNA for PCR (typically, order of the number of sampled cells), and good sequence reads. A single PCR cycle efficiency was estimated to equal $\epsilon=1.8$ \cite{Bolotin2012}, and {  there was} a typical number of about $30$ cycles { performed}. {  Viewing it as} a branching process {  one obtains an expression for variance on exit $\sigma_{p_i,PCR}^2\approx(2/\epsilon-1)\sigma_{p_i,cDNA}^2$ \cite{Grimmett2001}.} {  It yields} a factor $2/\epsilon=10/9$ correction to the variance of cDNA sampling.

{  If in question,} the estimator of the probability density distribution can be also obtained then:
\begin{equation}
\label{eq3}
W_c(p)=-\frac{d}{d p} \hat{F}_c(p).
\end{equation}

For each data set we estimate the upper $95\%$ confidence bound (further, CI95) for the frequency of unseen clonotypes (viewing observation failure as an outcome of binomial samplings) as 
\begin{equation}
\label{eq4}
p_0=1-(1-0.95)^s, \ s=\sum\limits_{k=1}^K\frac{1}{N_k},
\end{equation} 
{  where $s$ can be viewed as an effective inverse number of trials}.
{  Apparently,} about and below this value the clonal frequency distribution cannot be reliably estimated from the available data. 

To assess the potential impact of the additional sources and types of statistical sampling errors as well as laboratory errors that escaped post-processing we also estimated the clonotype frequency variances from three parallel samples from an autoimmune donor independently taken and sequenced $10$ month after HSCT \cite{Bolotin2012} and plugged them into the kernel distribution estimator (\ref{eq1}). The result did not show significant deviations from the one obtained with the sampling bottlenecks approach.

{  Resulting frequency distributions can be analyzed by various approaches. Since they clearly exhibited intervals of power law decay $\hat{F}_c(p)\propto p^\alpha$, we performed linear polynomial fits in the double-log scales employing a standard least square method, determining respective exponents and their CI$95$. 

To test applicability of the parametric methods, previously developed for assessing clonal size distributions \cite{Sepulveda2010}, we developed a Poisson abundance model under an assumption of power law clonal frequency distribution within a certain range.
In particular, one describes sampling distribution by a Multinomial law}

{ 
\begin{equation}
\label{PAM1}
P[\{m_i\}|D,\alpha]=\frac{D!}{(D-N_c)!\prod_{i=1}^{N_c} m_i} [f_\alpha(0)]^{D-N_c}\prod_{i=1}^{M}[f_\alpha(i)]^{m_i},
\end{equation}
where $D$ is the total diversity of the T cell repertoire, $m_i$ is the number of distinct TCR sequences found in $i$ copies, $M=\sum m_i$, $f_\alpha(i)$ is the parameter-dependent probability to obtain $i$ copies of a TCR variant in a sample. The latter can be calculated taking into account the typically small size of the sample comparing to the whole repertoire, which yields the Poissson distribution 
\begin{equation}
\label{PAM2}
f_\alpha(i)=\frac{e^{-p}p^i}{i!}
\end{equation}
for the clonotypes with identical frequency $p$. If the frequencies are expected to follow some distribution, in particular, $W(p)\propto p^{\alpha-1}$, integrating out $p$ gives 
\begin{equation}
\label{PAM3}
f_\alpha(i)\propto \frac{\Gamma(i+\alpha)}{i!}, \ i>-\alpha.
\end{equation}
Diversity $D$ and distribution power law exponent $\alpha$ are then numerically estimated by maximizing the (log)likelihood function (\ref{PAM1}). 
The method is easily adapted to exclude low and high clonal frequency outliers that do now follow power law: it suffice to redefine $f_\alpha(0)$ in (\ref{PAM1}) as the probability to fall off the the specified range of frequencies. 

} 

\section*{Results}

We assess clonal frequency distributions ascribing each distinct TCR$\beta$ CDR3 sequence to a distinct T cell clonotype. We make use of publicly available TCR libraries obtained from human donors as described in Materials \& Methods. Aiming to avoid prior assumptions on the clonal frequency distribution, we choose a non-parametric approach. As it is outlined in Materials \& Methods we construct an estimator based on a complementary cumulative normal distribution kernel. For each clonotype the Gaussian is centered at the measured frequency and has the variance calculated from the binomial sampling model as described in the above. Altogether, it allows to incorporate statistical sampling error, different for clonotypes with different frequencies, avoiding the blind cutoff of the less frequent ones.

\begin{figure}[!ht]
\begin{center}
\includegraphics[width=0.9\columnwidth,keepaspectratio,clip]{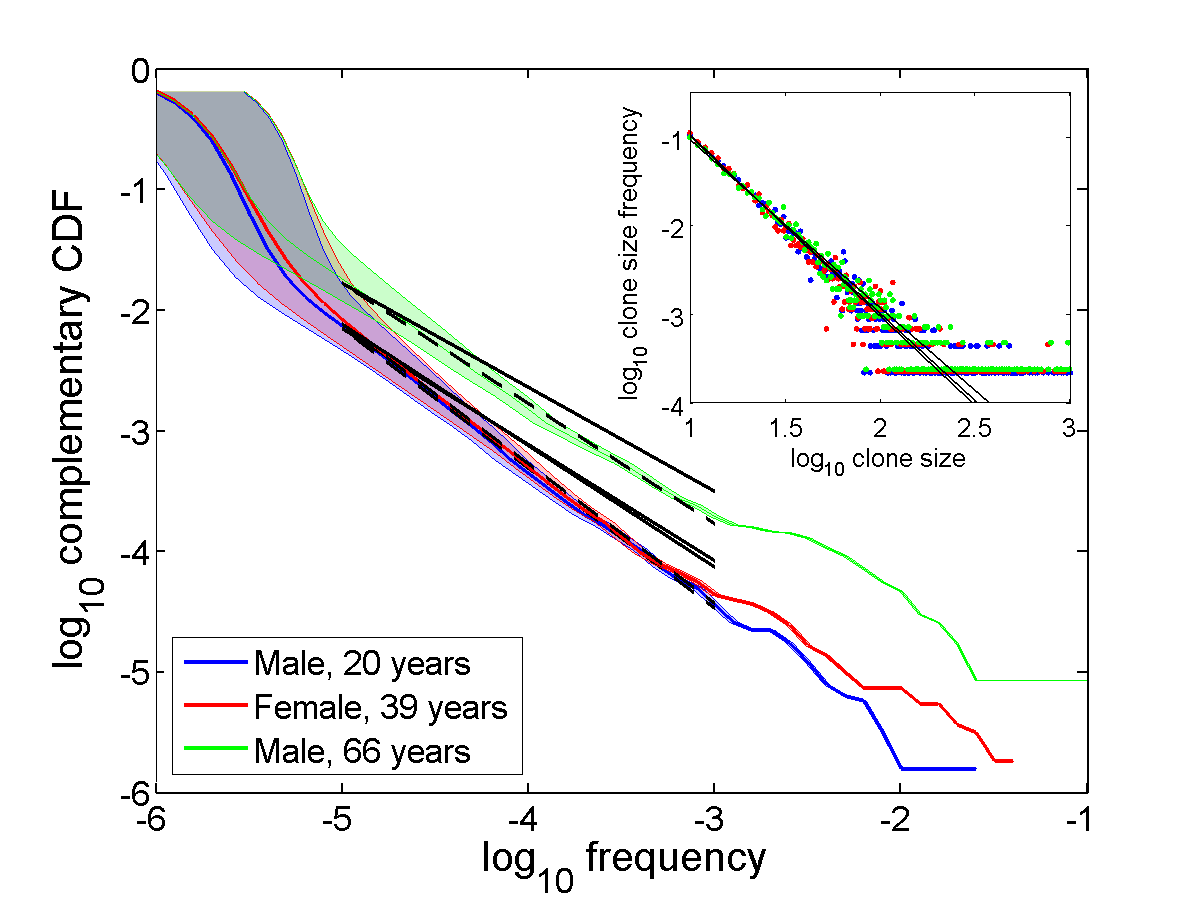}
\caption{{\bf Clonal statistics for healthy donors.} {  {\it Main figure.} Representative} complementary cumulative clonal frequency distributions $\hat{F}(p)$ (CDF) for three healthy individuals {  across different age groups} \cite{Britanova2014} in double log scale: {  male (20 years), female (39 years), and male (66 years)}. {  Shaded areas indicate CI$95$ intervals for clonal frequencies.} Black dashed lines show power law dependencies $\hat{F}(p)\propto p^\alpha$, $\alpha=-1.16$, $\alpha=-1.16$, and $\alpha=-1.0$, respectively, indicating a good fit of experimental data over two decades (cf. Fig.\ref{Figure2a} and Table \ref{Table2} for more details). {  Black solid lines show power law dependencies with the exponents derived from the Poisson abundance model fit (\ref{PAM1})-(\ref{PAM3}). {\it Inset.} Parametric approach \cite{Sepulveda2010}: frequency distribution of clonotypes binned by detected size in double log scale (colors code the same donors as in the main figure) and Poisson abundance method fits (black solid lines).}}
\label{Figure2}
\end{center}
\end{figure}

First, we analyze NGS TCR libraries obtained from reportedly healthy individuals:  {  $38$ subjects aged $9$ -- $90$ years \cite{Britanova2014}} and two male and one female, middle-aged (``Male 1'', ``Male 2'', and ``Female'') \cite{Warren2011}.
{  Representative examples for donors from different age groups} shown in Fig.\ref{Figure2}, {  illustrate the general result:} the major parts of the complementary cumulative frequency distributions $F(p) \equiv\mbox{Prob}(p_i\ge p)$ exhibit a power law decay over {  at least two} decades:
\begin{equation}
\label{eq6a}
\begin{aligned}
&\hat{F}(p)\propto p^{\alpha},\\
\end{aligned}
\end{equation}
where the exponent values are $\alpha=\{-1.16, -1.16, -1.0 \}$. {  Noteworthy, deviations from the power law are observed not only in the low frequency range (which can be accounted for significant statistical sampling uncertainties), but for abundant clonotypes as well. The latter hints that the clonal size distribution might consist of at least two different parts. 

Parametric approach that makes use of Poisson abundance models \cite{Sepulveda2010} also produces good results, once a power law distribution is assumed (Fig.\ref{Figure2}, inset). However, turning from probability density to cumulative distributions, one notices moderate but systematic underestimation of the exponent value (Fig.\ref{Figure2}, main). A possible reason for that could lie in employing probability estimators, more sensitive to statistical sampling errors than cumulative ones, used in the developed non-parametric method.}

\begin{figure}[!ht]
\begin{center}
\includegraphics[width=0.9\columnwidth,keepaspectratio,clip]{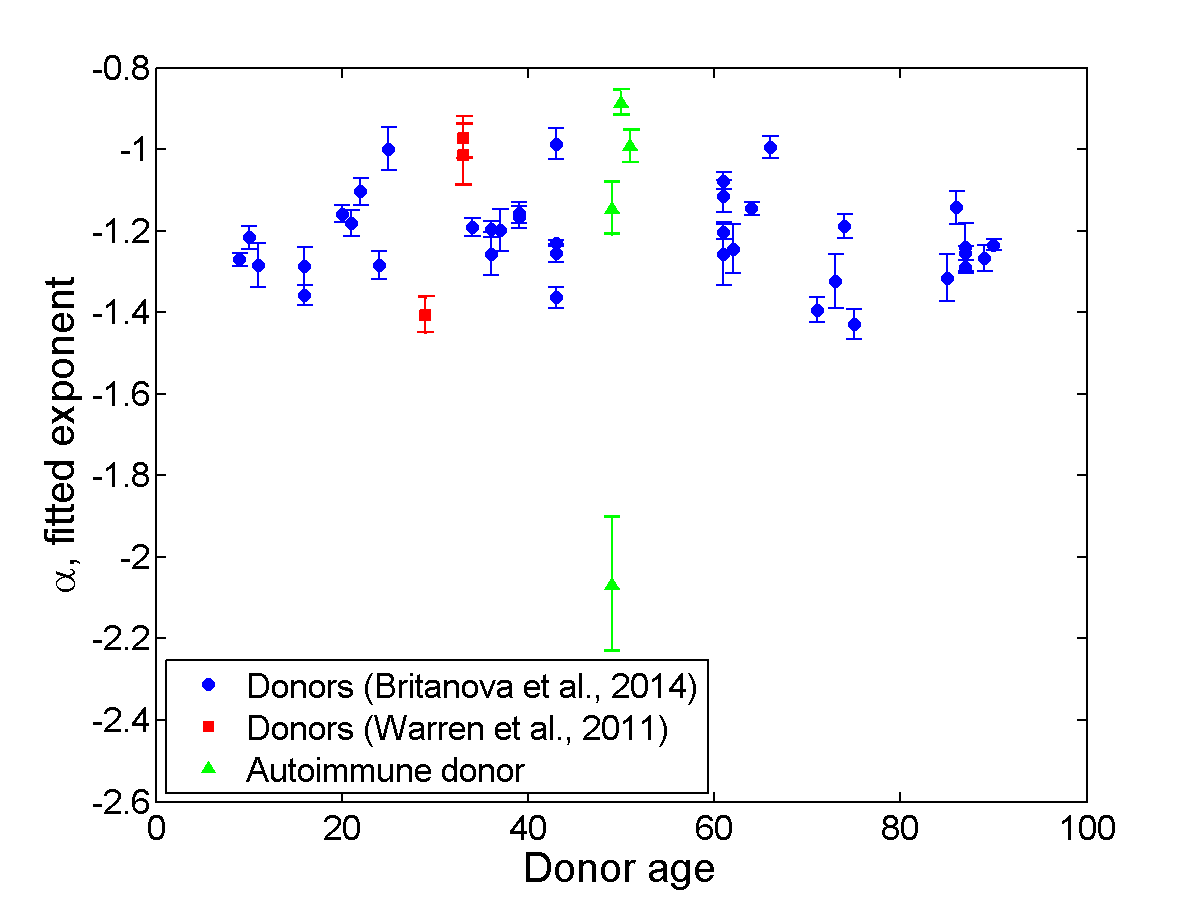}
\caption{{\bf Power law exponents.} {  Exponents $\alpha$ of power law fits with respective CI$95$ indicated vs. age of individuals. Blue circles: healthy donors from \cite{Britanova2014}, least square fits performed over the interval $\log_{10}p\in[-5.5, -3.5]$ or $\log_{10}p\in[-5.0, -3.0]$, whichever produced better quality. Red squares: healthy donors from \cite{Warren2011}, least square fits performed over the interval $\log_{10}p\in[-4, -2]$. Green triangles: autoimmune patient before and after treatment \cite{Mamedov2011,Bolotin2012,Britanova2012}, least square fits performed over the interval $\log_{10}p\in[-3.5, -2.5]$ for the time point before treatments and $\log_{10}p\in[-4, -2]$ for the three time points after.}}
\label{Figure2a}
\end{center}
\end{figure}

{  The fitted exponents along with respective CI$95$ are shown in Fig.\ref{Figure2a} vs. donors age (see also Table \ref{Table2} for details). In all cases they were obtained over two decades of clonal frequencies, specific ranges being $\log_{10}p\in[-5.5,-3.5]$ or $\log_{10}p\in[-5,-3]$ (whichever produces a better fit) for the data from \cite{Britanova2014}, and $\log_{10}p\in[-4,-2]$ for the data from \cite{Warren2011}. Overall, the power law exponents fall into quite a narrow range $\alpha\in[-1.43 -0.97]$. Notably, no pronounced age dependence is observed. One can only point it out that all values among the young ($9-20$ years) and elderly ($70-90$ years) groups belong to the lower half of the interval: $\alpha<-1.18$, though it could simply be due to moderate pool of donors.}

Secondly, we follow the dynamics of the clonal frequency distribution from an autoimmune patient(``Male A'') who undergone autologous HSCT and demonstrated a stable remission since \cite{Mamedov2011,Bolotin2012,Britanova2012}. Before the HSCT the patient's TCR repertoire had been analyzed for 2 years, and a number of stably hyperexpanded clones associated with inflammatory response were found.  NGS was performed right before HSCT (the reference $0$ time point), $4$, $10$, and $25$ months after the procedure. Reportedly, the treatment led to a major resetting of the repertoire with only about $10\%$ of earlier observed clonotypes detected afterwards. Previously hyperexpanded clones (including a pro-inflammatory one) drastically decreased their frequencies, while the other clones grew large and remained so over the study \cite{Britanova2012}. Interestingly, the share of hyperexpanded clonotypes in the repertoire even increased after HSCT, when remission reportedly occured, indicating that their existence is not a trait of an abnormal state.   

\begin{figure}[!ht]
\begin{center}
\includegraphics[width=0.9\columnwidth,keepaspectratio,clip]{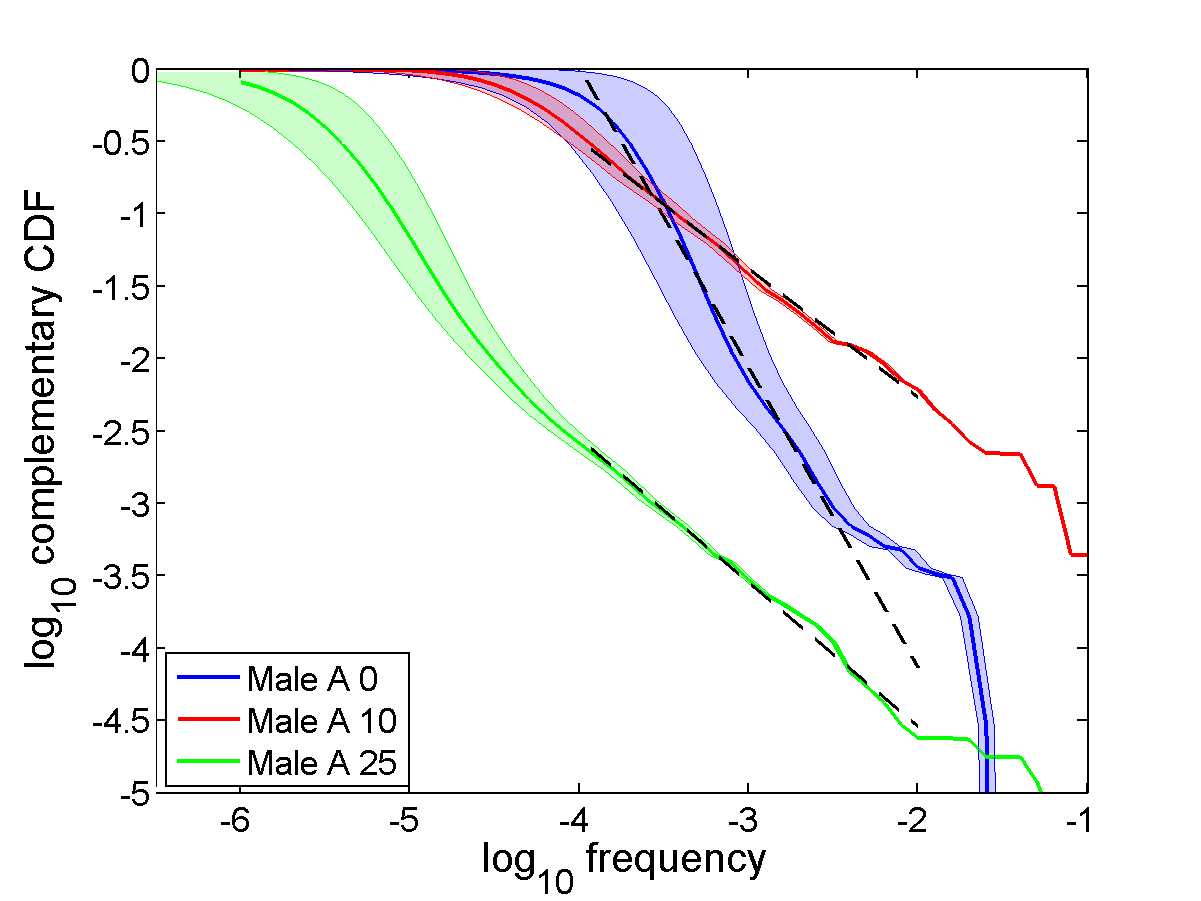}
\caption{{\bf Clonal statistics for an autoimmune patient.} Complementary cumulative clonal frequency distributions $\hat{F}(p)$ (CDF) for an autoimmune patient \cite{Mamedov2011,Bolotin2012,Britanova2012} in double log scale: right before the treatment (blue), 10 months after (red), and 25 months after (green). {  Shaded areas indicate CI$95$ intervals for clonal frequencies}. Dashed lines show power law fits $\hat{F}(p)\propto p^\alpha$, $\alpha=-2.07$, $\alpha=-0.88$  and $\alpha=-0.99$, respectively. {  Least square fits performed over the interval $\log_{10}p\in[-3.5, -2.5]$ for the time point before treatments and $\log_{10}p\in[-4, -2]$ for the time points after.}}
\label{Figure3}
\end{center}
\end{figure}

Reconstruction of frequency distributions before and after the treatment reveals a drastic change in the power law exponent from $\alpha\approx -2.07$ before HSCT to $\alpha\approx-0.99$ after $25$ months, though the power law fit of the major part of the distribution (spanning across two decades {  except for the horizontal axis for time point $0$}) remains plausible at all times (Fig.\ref{Figure3}, Table \ref{Table2}).  Remarkably, before HSCT the exponent was considerably below those obtained in reportedly healthy adult individuals {  (Fig.\ref{Figure2a})}. Conversely, at all time points after HSCT the exponent consistently remained in the interval, specific for healthy donors. 

Assessing the possible effect of technical differences in NGS procedures on the clonal frequency distribution, we note the following. {  The values obtained from healthy donors libraries in \cite{Warren2011} and \cite{Britanova2014} agree well.} Qualitatively different values of the power law exponent $\alpha$ are estimated from the TCR libraries assembled before, $4$ and $10$ months after HSCT from the same autoimmune donor by the same NGS protocol \cite{Mamedov2011}. Moreover, the values of $\alpha$ $4-25$ months after the therapy are typical of those for healthy donors, obtained by the protocols different in details. Therefore, differences in protocols do not seem to bias the value of the power law exponent significantly.

\section*{Discussion}

Next generation sequencing tools have recently made a high-throughput analysis of T cell repertoire possible. Through the last several years most attention has been paid to estimating clonal diversity, and the reported numbers of distinct clonotypes have been renewing record each time the methods improved and analysis deepened. At the same time the statistics of clonal frequency distributions remained a secondary issue, addressed mostly to estimate the number of ``unseen'' small-frequency clonotypes by Fisher's techniques \cite{Robins2009,Fisher1943}.

About ten years ago even very scarce data led to hypothesize that human T cell clonal frequency distributions have heavy non-Gaussian tails that could follow a power law \cite{Naumov2003}. However, the study was confined to only $141$ clonotype and allowed to produce a power-law-like dependence over a single decade of magnitude at best, rendering the statistical validity of observations and fitted exponents questionable, according to the general practices \cite{Stumpf2012}. A recent study by Meier et al. \cite{Meier2013} generated much deeper libraries, reportedly containing about $10^5 - 10^6$ distinct TCR sequences. They also shared the much anticipated view of the self-similar properties of the repertoire, but admitted several shortcomings in their quantitative analysis. First, as we already mentioned in Introduction, the authors dismissed clonotypes with frequencies below $0.05\%$, restricting themselves to a single order of magnitude interval, insufficient for a reliable power law fit \cite{Stumpf2012}. Second, they estimated clonal frequency distributions with unevenly sized bins, which distorted the result. Another recent report studied evolution of clonal distributions in the course of HSCT, also producing evidence of heavy non-Gaussian tails, but did not attempt to quantify it \cite{Heijst2013}. To achieve a progress one had to employ appropriate statistical inference techniques, taking into account sampling errors of sequencing but avoiding an overly strong data cutoff. 

Parametric fitting approach offered several model distributions, which became popular choices in analyzing restricted TCR repertoires from mice, to name uni- and multi-variate Poisson abundance models with log-normal, exponential, and gamma sampling rates \cite{Sepulveda2010,Rempala2011}. However, as it was demonstrated in \cite{Sepulveda2010}, even a systematic analysis of the fit goodness cannot not yield a decisive answer in favor of one or another model, when experimental data are confined to about a hundred of clonotype frequencies. Besides, parametric estimators have two obvious limitations for analyzing clonal statistics {\it per se}. First, it requires to assume a specific functional distribution of TCR frequencies. Its biological meaning is unclear since there is no bottom-up theory predicting such a distribution from first immunological principles. Moreover, assuming a model distribution we limit our ability to infer these first principles in the top-bottom analysis. Second, it is not guaranteed that a single model distribution describes the whole range of frequencies, and optimal model distributions could, in principle, differ between species or even donors. {  On the other hand, when the kind of functional dependence(s) and their validity intervals are known {\it a priori}, by the virtue of another analysis, parametric fitting becomes indispensable.}

{  To overcome these difficulties} we proposed to employ non-parametric analysis, appropriate for the currently available extensive human TCR libraries. We have successfully implemented kernel density estimators, which conveniently incorporate the variances of the clonal frequency estimators, essentially dependent on the clonotype size. Variances can either be calculated by comparing data from several parallel samples taken and processed independently (which is preferable, as it counts both statistical and experimental errors), or by estimating statistical sampling errors at the bottlenecks of an experiment. One has to keep in mind {  that non-parametric fitting cannot overcome} limitations of NGS as the data source {  sequencing errors, PCR amplification bias, etc.} Meanwhile, the developed error detection and correction algorithms {  along with the most recent cDNA barcoding technique \cite{Britanova2014}} already seem to ensure substantial reliability (for a detailed discussion of the issue we refer a reader to \cite{Bolotin2012,Venturi2012}). {  It is worth stressing that non-parametric analysis does not dismiss parametric one, as the latter is required to evaluate and validate the functional dependence in the output data from the former.}   

Implementing this approach, we demonstrated for the first time that cumulative clonal frequencies distributions from {  41} adult human donors can be fitted well with the power law $F(p)\propto p^{\alpha}$ over several decades of magnitude. For reportedly healthy individuals the exponents grouped about $\alpha\approx-1...-1.4$, while for an autoimmune donor it had been considerably less, about $\alpha\approx-2.1$ before HSCT and increased to the values typical of healthy donors $\alpha\approx-1$ afterwards. We were unable to identify any certain functional behavior of the frequency distribution for small size and hyper-expanded clonotypes, in the first place, due to insufficient statistics in current experimental data: the former yield reads of very low copy number or simply escape sequencing, the latter are too few. We, therefore, cannot exclude deviations from the power law scaling in these parts of distributions. 

These findings put a challenge for theoretical immunology to identify the mechanism(s) behind such clonal distributions. Several candidates appear plausible. A major role in shaping individual clonal sizes belongs to T cell competition for access to cognate p-MHCs on antigen presenting cells, for survival and proliferation stimuli. Indeed, it is known that the same antigen can be recognized by several, sometimes hundreds of clonotypes \cite{Pacholczyk2006,Kedzierska2006}, and, vice versa, a given TCR can recognize theoretically up to $10^6$ different p-MHCs due to inherent cross-reactivity \cite{Mason1998,Baker2012}. A good question to address is whether existing mathematical models of T cell competition and clonal selection at the periphery \cite{deBoer1997,Callard2003,Stirk2008,Stirk2010b,Stirk2010a,Ivanchenko2011} can reproduce power-law distributions, or their further development is needed. Another crucial mechanism is the positive and negative selection in thymus, and we await theoretical and experimental advances here. Some insight could also be expected from the statistics of V(D)J recombination, though it still remains a far stretch to infer the functional properties from TCR sequence.

Finally, our results should inspire studying a greater number of human TCR libraries to confirm or disprove that statistical T-cell clonal size distributions for healthy and autoimmune donors (at least for certain diseases) may exhibit drastically different power laws. Another issue to be addressed in a consistent manner is that clonal frequency distribution may possibly evolve significantly under treatment like HSCT, similarly to drastic changes in the other numerical measures \cite{Meier2013,Heijst2013}. We believe that the proposed approach will become a useful tool in the studies of immunity and autoimmunity development, complement the developing deep sequencing methods of individual diagnostics of infectious and autoimmune diseases, characterizing and understanding immunosenescence.

\section*{Acknowledgements}
The Authors thank D.M. Chudakov, D.A. Boilotin, Ya. Safonova, and J. Carneiro for illuminating discussions, and also the Max Planck Institute for the Physics of Complex Systems for providing stimulating scientific environment and hospitality.

\bibliography{Ivanchenko_pone}

\section*{Figure Legends}

\clearpage

\section*{Tables}

\begin{table}[!ht]
\caption{
\bf{Characteristics of the TCRB libraries \label{Table1}}}
{\begin{tabular}{llllllll}\hline
Subject, age & Timepoint & No. of  & Sequences & Distinct & Average & Unseen clonotype & Maximal \\
 & (month) & T cells & read & reads & frequency & frequency CI95 & frequency \\
 & & ($\times10^6$) & ($\times10^{6}$) & ($\times10^3$) & ($\times10^{-4}$) & ($\times10^{-4}$) & ($\times10^{-2}$)\\\hline
Male, 29 & -- & 12.0 & 188 & 494 & 0.0002 & 0.8 & 0.1 \\
Male, 33 & -- &  16.4 & 6.2 & 193 & 0.0005 & 1.0 & 0.2 \\
Female,33 & -- & 18.2 & 1.1 & 94 & 0.001 & 3.0 & 0.6 \\
\hline
Group 1 ($9-25$ y) & -- & 3.0 & 0.846 & 490 & 0.0058 & 0.037 & 0.25  \\
Group 2 ($36-43$ y) & -- & 3.0 & 0.948 & 446 & 0.0047 & 0.0331 & 7.6 \\
Group 3 ($61-66$ y) & -- & 3.0 & 0.999 & 308 & 0.0031 & 0.03 & 8.6 \\
Group 4 ($71-90$ y) & -- & 3.0 & 0.995 & 354 & 0.0035 & 0.03 & 7.0 \\
\hline
Male A, 49 & 0 & 2 & 0.011 & 6.1 & 1.5 & 0.026 & 2.1  \\
Male A, 49 & 4 & 2 & 0.014 & 3.0 & 3.3 & 0.021 & 9.1 \\
Male A, 50 & 10 & 3.5 & 1.07 & 2.3 & 4.4 & 0.0015 & 19.5 \\
Male A, 51 & 25 & 3.5 & 1.06 & 168 & 0.59 & 0.0015 & 9.4 \\
\hline
\end{tabular}}{}\\
{Sources of libraries: upper block \cite{Warren2011}, middle block \cite{Britanova2014}, bottom block \cite{Mamedov2011,Bolotin2012,Britanova2012}. Values in the middle block are averages within each group.}
\end{table}

\begin{table}[!ht]
\caption{
\bf{Power law exponent fits for complementary cumulative clonal frequency distributions and $95\%$ confidence intervals. \label{Table2}}}
{\begin{tabular}{llll}\hline
Subject, age & Timepoint & $\alpha$  & \ CI$95_\alpha$ interval \\
 & (month) &  &  \ width \\\hline
Male, 29 & -- & -1.41 & \ 0.17 \\
Male, 33 & -- & -0.97 & \ 0.10 \\
Female, 33 & -- & -1.01 & \ 0.06 \\\hline
Group 1 ($9-25$ y) & -- & -1.21 & \ 0.07  \\
Group 2 ($36-43$ y) & -- & -1.20 & \ 0.06 \\
Group 3 ($61-66$ y) & -- & -1.15 & \ 0.07 \\
Group 4 ($71-90$ y) & -- & -1.28 & \ 0.08 \\
\hline
Male A, 49 & 0 & -2.07 &  \ 0.33 \\
Male A, 49 & 4 & -1.15 & \  0.12 \\
Male A, 50 & 10 & -0.89 & \  0.57 \\
Male A, 51 & 25 & -0.99 &  \ 0.58 \\
\hline
\end{tabular}}\\
{Sources of libraries: upper block \cite{Warren2011}, middle block \cite{Britanova2014}, bottom block \cite{Mamedov2011,Bolotin2012,Britanova2012}. Values in the middle block are averages within each group.}
\end{table}

\end{document}